%% file: acl_latex.tex
\documentclass[11pt]{article}

\usepackage[final]{acl}

\usepackage{times}
\usepackage{latexsym}

\usepackage[T1]{fontenc}

\usepackage[utf8]{inputenc}

\usepackage{microtype}

\usepackage{inconsolata}

\usepackage{graphicx}

\usepackage{amssymb}
\usepackage{xcolor}
\usepackage{booktabs}
\usepackage{multirow}
\usepackage{tikz}
\usetikzlibrary{positioning, arrows.meta, fit, calc, backgrounds}
\usepackage{enumitem}
\setlist{nosep, leftmargin=*}
\usepackage{cleveref}
\usepackage{xurl}  

\title{\textsc{RECAP}: An End-to-End Platform for Capturing, Replaying,\\and Analyzing AI-Assisted Programming Interactions}

\author{Keyu He* \quad Qianou Ma* \quad Valerie Chen \quad Wayne Chi \quad Tongshuang Wu\\ 
Carnegie Mellon University\\
\texttt{\{keyuhe, qianouma\}@cmu.edu}}

\begin{document}
\maketitle

\begin{abstract}
  Understanding how developers interact with AI coding assistants requires more than chat logs or git histories in isolation; it requires reconstructing the full context: which prompt led to which edit, what the developer tried and discarded, and how their strategy evolved over time.
  We present \textsc{RECAP} (\textbf{R}eplay and \textbf{E}xamine \textbf{C}aptured \textbf{A}I \textbf{P}rogramming), an open-source platform that (1)~passively records AI chat sessions and fine-grained code edits inside VS~Code without disrupting the developer's workflow, (2)~merges them into a unified timeline for interactive session replay, and (3)~exposes an extensible analysis layer, with example modules for behavioral classification and AI reliance measurement.
  Deployed in a university software engineering course, \textsc{RECAP} captured 2,034 prompts and 8,239 code edits from 41 students across a multi-week project.
  We demonstrate how the platform's linked data and replay capabilities enable analyses of developer-AI interaction patterns that no single data source could support.
  \textsc{RECAP} is available on the VS~Code Marketplace.\footnote{\label{footnote:link}\url{https://marketplace.visualstudio.com/items?itemName=Copilot-Archiver.copilot-archiver}}
\end{abstract}

\section{Introduction}

AI coding assistants such as GitHub Copilot, Cursor, and ChatGPT are now part of everyday programming \cite{Dohmke2023-nd}.
As adoption accelerates, researchers and educators need to understand \emph{how} developers actually use these tools and \textit{what} impact do users' behavioral patterns have on the code produced.
Answering these questions requires \emph{reconstructing the full interaction context}: which prompt led to which edit, what the developer saw when they accepted or discarded a suggestion, and how their strategy evolved across a project. This is increasingly difficult as AI coding tools become more \emph{agentic}: a single prompt may span multiple files, invoke tools (search, terminal, test runners), and iterate over many turns before producing a result. Two challenges follow for instrumentation. (i) \textbf{Linkage:} chat logs and git histories examined in isolation lose the causal link between a prompt and the edits it produced. (ii) \textbf{Long horizon:} sessions span hours or days, well beyond the minutes of typical lab studies.

Prior studies of AI-assisted programming span lab-based usability experiments~\cite{barke2023grounded,vaithilingam2022expectation}, real-world log collections on code completions~\cite{chi2025copilotarenaplatformcode}, classroom deployments with custom interfaces~\cite{kazemitabaar2024codeaid,babe2024studenteval}, notebook-only environments \cite{ma2026dspm}, or shorter, simpler tasks \cite{Zhang2026-je}.
These works do not capture the complexities of modern agentic coding, which involves multi-file, long-horizon workflows; there remains a need for naturalistic, replayable, and scalable instrumentation that links prompts, suggestions, and fine-grained code edits across extended development sessions.

To address the gap, we present \textsc{RECAP},\footnote{Demo Video: \url{https://www.youtube.com/playlist?list=PLkTxDosSc5HnD1cxi0e_aGRvPZx4ZCfdv}} a platform designed for researchers, CS educators, and tool builders to observe and analyze AI-assisted programming in its natural setting. \textsc{RECAP} has two core components, supported by an extensible analysis layer:

\begin{enumerate}
  \item \textbf{Copilot Interaction Archiver}: a VS~Code extension that passively captures AI chat sessions and a fine-grained shadow git history of every code change, then uploads them to cloud storage with privacy-preserving hashing.

  \item \textbf{Session Replay Viewer}: a web application that reconstructs the developer's full interaction context by merging chat and code streams into a unified chronological timeline, enabling researchers to step through a session and see exactly which prompt led to which code change.
\end{enumerate}

On top of the timeline, we provide example analyses (behavior classification, AI reliance attribution, prompt embeddings) that demonstrate what the platform enables; researchers can plug in their own without modifying the capture layer.

\input{figures/architecture}

We deployed \textsc{RECAP} in a university applied machine learning course where 41 students used GitHub Copilot on a two-week project.
The system captured 2,034 prompts and 8,239 code edits.
We present this deployment as a demonstration of what the platform enables, not as a standalone empirical study; the analyses below are exploratory illustrations of what the platform enables.
\textsc{RECAP} is open-source and released under the MIT license.

\section{System Architecture}

\subsection{Design Rationale}

Researchers studying AI-assisted programming want to answer questions such as: How do developers' AI usage strategies evolve over multi-week projects? What fraction of AI-suggested code survives into the final product? How does reliance on AI differ across task types or experience levels?
Answering these questions requires two capabilities that existing tools do not provide together.

The first is \emph{capturing fine-grained code edits over extended, real-world projects}. Prior recording tools target short, controlled tasks; standard version control lacks the temporal resolution needed for longer efforts, where a typical git commit may bundle hours of work, including dozens of prompts, accepted and rejected suggestions, manual edits, and debugging attempts.
The second is \emph{linking AI prompts to code edits}. Chat logs and code histories are recorded in separate systems with no shared identifiers: a conversation transcript says ``I inserted code into file X,'' but the git history has no record of which commit corresponds to that insertion.

\textsc{RECAP}'s architecture addresses both challenges.
For fine-grained capture, a shadow git repository records a commit on every file save and even on unsaved in-editor changes, providing the temporal resolution needed to isolate individual edits throughout a project.
For prompt-to-edit linking, AI chat responses include \emph{text edit groups} (TEGs)---the exact file paths and content the AI proposed to insert.
By matching TEGs against subsequent shadow git diffs within a temporal window using fuzzy line-level comparison, the pipeline attributes each code edit to a specific AI response, a human edit, or an external source.
Together, these two mechanisms enable both the interactive replay and the downstream analyses.
Figure~\ref{fig:architecture} shows how the two core components, data collection and replay platform, connect through this shared timeline, with extensible analyses built on top.
The following subsections describe each component: the Copilot Interaction Archiver (\S\ref{sec:collection}) and the Session Replay Viewer (\S\ref{sec:replay}), followed by example analyses (\S\ref{sec:analysis}).

\subsection{Copilot Interaction Archiver}
\label{sec:collection}

The {Copilot Interaction Archiver} is a VS~Code extension that captures two primary data streams, chat sessions and code edits, without disrupting the developer's workflow.
The two streams are what make linking possible: each is timestamped, allowing the analysis pipeline to reconstruct which prompts preceded which edits.

\paragraph{Chat sessions.}
The extension watches VS~Code's workspace storage, where GitHub Copilot persists each conversation as a UUID-named JSON file.
When a session file is created or modified, the extension reads the full conversation, which includes user prompts, AI responses, tool calls, code references, and model metadata, and uploads it with a 10-second debounce.
The JSON includes \textbf{text edit groups} (TEGs): the exact file paths and content that the AI proposed to insert.
TEGs are what allow the replay viewer to attribute specific code edits to specific AI responses.

\paragraph{Workspace code edits.}
A hidden git repository (\texttt{.archiver\_shadow/}) mirrors the user's workspace.
On every file save, create, delete, or rename, the extension copies the affected file into the shadow repo and commits it with a labeled message indicating the operation type.
Unsaved in-editor changes are also captured as \textsc{dirty snapshot} commits (5-second debounce, 30-second maximum interval), preserving even discarded edits.
This provides a diff-able history far more fine-grained than the developer's own git commits, which may bundle hours of work into a single commit.
The shadow repo is synced to cloud storage as a git bundle on a 5-minute cooldown.

\paragraph{Privacy.}
User identifiers are SHA-256 hashed client-side before any network request.
The backend (Express.js with JWT authentication) forcefully prefixes upload paths with the authenticated user's hash, preventing path traversal.
The extension never holds cloud credentials; it requests short-lived presigned URLs for each upload.

\begin{figure*}[t]
  \centering
  \includegraphics[width=\textwidth]{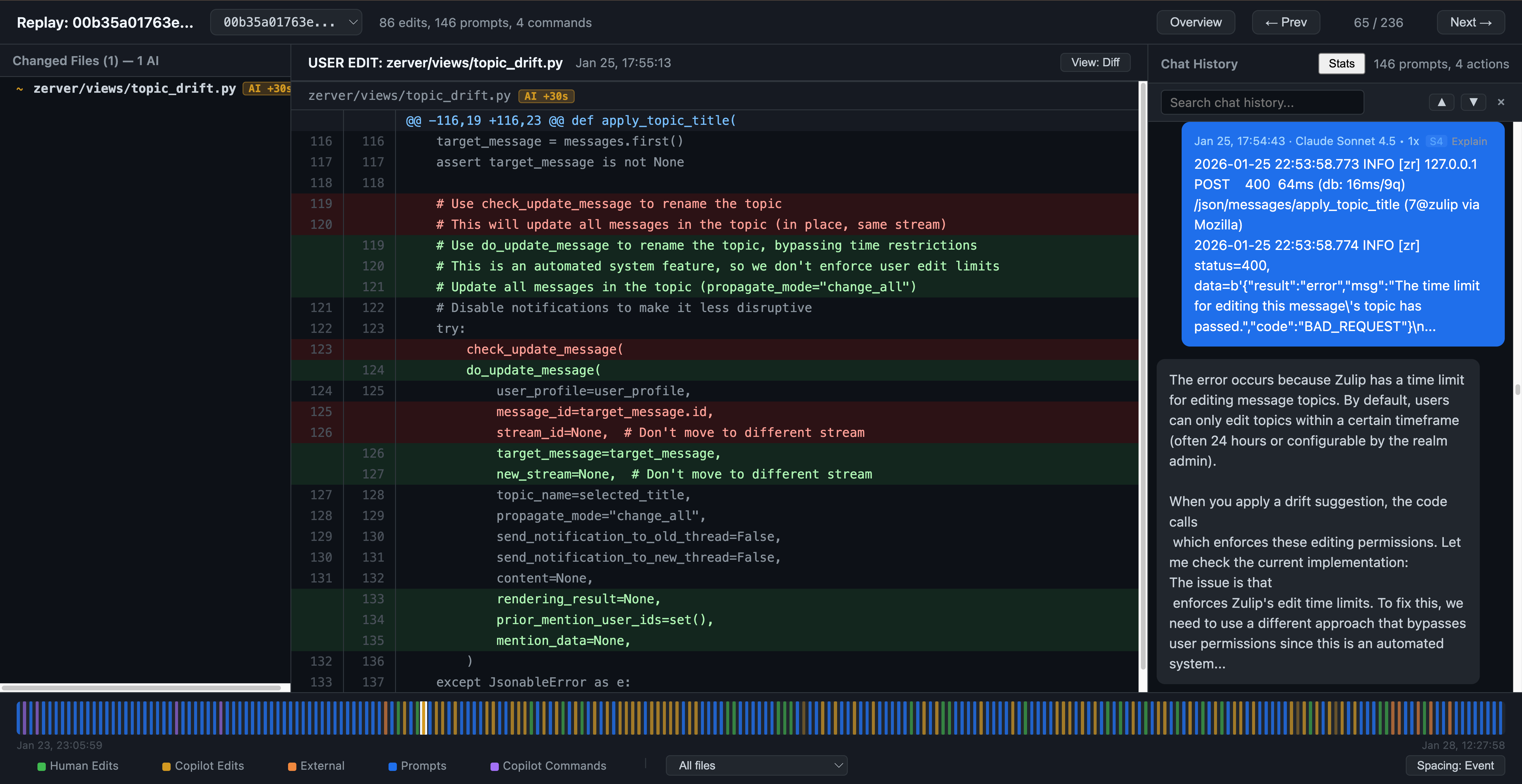}
  \caption{\textbf{\textsc{RECAP} Replay Viewer.} Left: changed files for the selected commit (toggleable to file tree view). Center: unified diff view showing a code change attributed to Copilot. Right: chat panel with the corresponding AI conversation. Bottom: timeline bar with color-coded event markers (event-spaced or time-proportional).}
  \label{fig:replay}
\end{figure*}

\subsection{Session Replay Viewer}
\label{sec:replay}

The Replay Viewer is a self-contained web application that makes the collected data browsable and analyzable (Figure~\ref{fig:replay}).
It merges shadow git commits and chat events into a single chronological timeline, allowing a researcher to step through a session. 
For example, the researcher can see that a student asked the AI to implement a feature, the AI proposed code changes across multiple files, and subsequent commits show the student accepting some suggestions while rewriting others.

\paragraph{Interface.}
The viewer presents a four-panel layout:
(1)~a \textbf{file tree} showing the workspace at the selected commit, with change indicators and AI-attribution badges;
(2)~a center \textbf{diff view} rendering GitHub-style unified diffs with per-file AI attribution;
(3)~a \textbf{chat panel} displaying all chat sessions merged chronologically, with the active message highlighted; and
(4)~a \textbf{timeline bar} at the bottom with color-coded markers (green for human edits, yellow for Copilot edits, orange for suspected external sources, blue for chat prompts, purple for agent actions) and keyboard navigation.
The viewer supports file-based filtering (double-click a file to see only commits touching it), searchable chat history, and time-proportional or event-spaced timeline modes.

\paragraph{Edit attribution.}
For each git commit, the pipeline determines \emph{who wrote the code} by matching text edit groups (TEGs) from Copilot's chat responses to subsequent git diffs within a 5-minute window.
Because the TEG representation in the chat JSON may differ from the committed code in formatting, the matching uses fuzzy line-level comparison, yielding a per-file match score.
Edits are classified as full matches, partial matches (typically due to formatting differences or developer modifications to the suggested code), or unmatched.
For unmatched commits, a separate heuristic flags edits as likely from an external source if the net new content exceeds a size threshold or the implied typing speed exceeds 100~WPM.
Clicking an AI badge in the viewer reveals the matched TEG, its source prompt, the match score, and the time delta between the AI response and the commit.

\begin{figure}[t]
  \centering
  \includegraphics[width=\columnwidth]{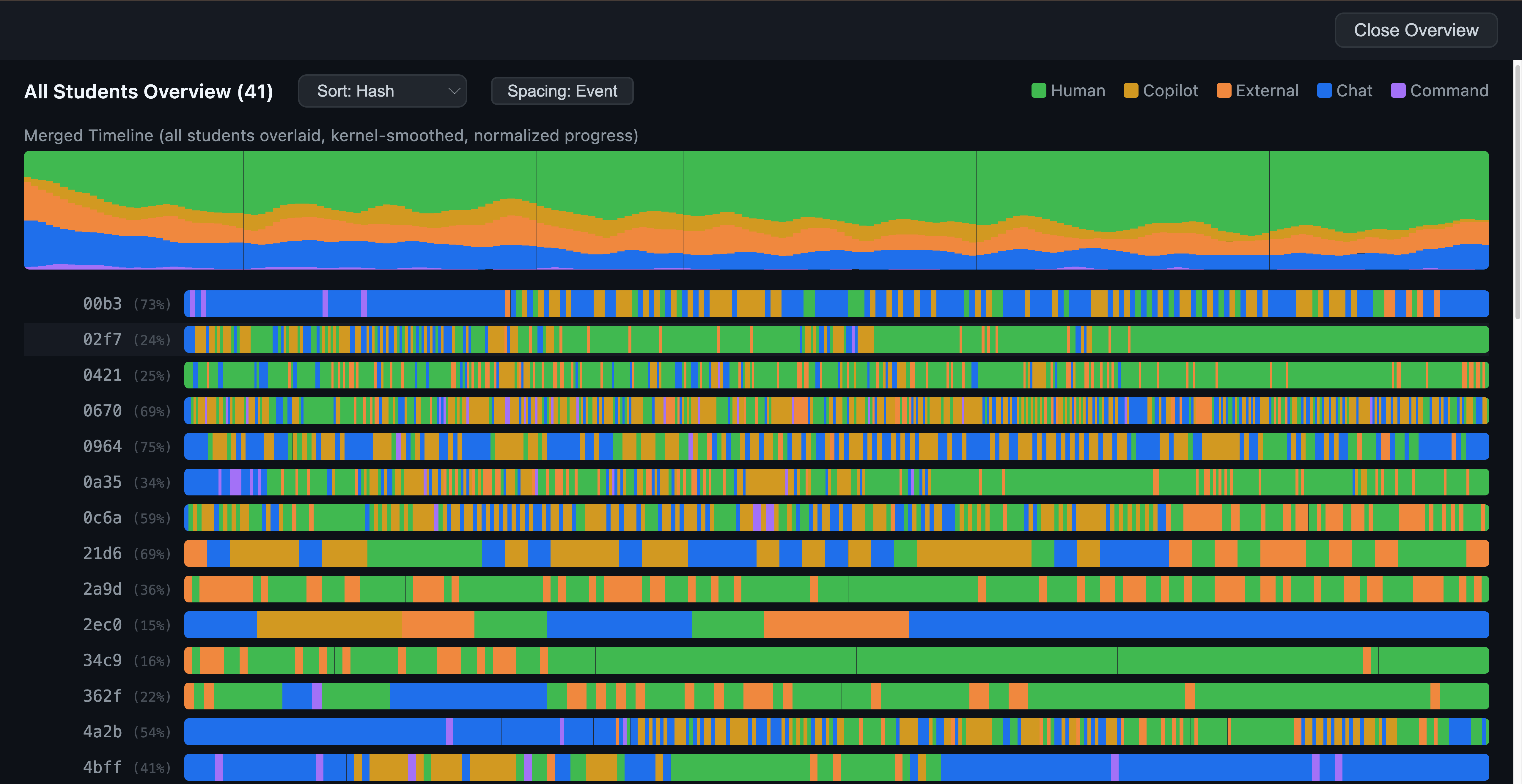}
  \caption{Multi-student overview panel. Each row is one student's event-spaced timeline. Colors: green = human edits, orange = Copilot, blue = chat prompts. Top: merged density across all students.}
  \label{fig:overview}
\end{figure}

\paragraph{Multi-student overview.}
In classroom deployments, the viewer provides an overview panel displaying all students' timelines sorted by AI edit share, with a merged density visualization showing the aggregate distribution of event types over normalized project progress.
Instructors can quickly identify outliers---students with unusually high AI reliance or irregular work patterns---and click through to inspect individual sessions (Figure~\ref{fig:overview}).

\paragraph{Offline mode.}
The viewer also supports drag-and-drop loading of exported timeline files for fully offline analysis without a server.

\subsection{Example Analyses}
\label{sec:analysis}

\textsc{RECAP} includes analysis modules that demonstrate what the linked data enables.
These serve as starting points; the pipeline is designed for researchers to extend with their own analyses.

\paragraph{Behavior classification.}
Each prompt is classified using an LLM with a codebook of 17~behavior codes in 6~categories (Table~\ref{tab:codebook}). Four categories---Plan, Code, Explain, and Eval---are adopted from \citet{ma2026dspm}. We found that a substantial share of prompts in our deployment fell outside those categories, so we introduced two additional categories: \textbf{Setup} (environment configuration, git operations, and deployment) and \textbf{Converse} (acknowledgments, greetings, and context sharing).

\paragraph{AI reliance metrics.}
Timeline events are segmented into work sessions (30-minute inactivity gap), then human, Copilot, and external edits are counted per session, yielding a per-session AI edit share that can be tracked over time.

\paragraph{Prompt embedding and clustering.}
Prompts are embedded using a multilingual sentence transformer \cite{reimers-2019-sentence-bert}, projected to 2D via t-SNE \cite{Maaten2008-dm}, and clustered with KMeans (inspirations drawn from Hodoscope \cite{zhong2026hodoscope}). 
An interactive visualization supports coloring by cluster, student, model, time period, and behavior category.

\paragraph{Extensibility.}
Beyond these modules, the data collection extension already captures supplementary streams: a paste watcher logs clipboard pastes and large insertions (used by the ``external AI'' heuristic), and a log watcher tails the Copilot debug log for agent intent and completion events.
The chat session JSON also includes tool calls with terminal commands and exit codes, chain-of-thought reasoning traces, and TODO lists generated by the agent.
The architecture supports adding new telemetry sources, such as window focus events and cursor tracking.

\begin{table}[t]
  \centering
  \resizebox{\linewidth}{!}{%
    \begin{tabular}{@{}llp{4.5cm}@{}}
      \toprule
      \textbf{Category}         & \textbf{Code}                         & \textbf{Description}                         \\
      \midrule
      \multirow{4}{*}{Plan}     & \texttt{ai\_suggest\_steps\_or\_plan} & Step-by-step workflow or plan                \\
                                & \texttt{ai\_breakdown\_intent}        & Decompose complex goal                       \\
                                & \texttt{ai\_improve\_prompt}          & Refine prompt wording                        \\
                                & \texttt{ai\_choose\_approach}         & Choose library, technology, or design        \\
      \midrule
      \multirow{3}{*}{Code}     & \texttt{ai\_generate\_code}           & Produce code for a requested action          \\
                                & \texttt{ai\_edit\_partial\_code}      & Edit a specific snippet or function          \\
                                & \texttt{ai\_write\_documentation}     & Write or edit docs, comments, READMEs, text            \\
      \midrule
      \multirow{4}{*}{Explain}  & \texttt{ai\_explain\_bug\_or\_error}  & Explain error or traceback and outline fix   \\
                                & \texttt{ai\_explain\_code\_or\_api}   & Interpret specific code or explain a function/API     \\
                                & \texttt{ai\_explain\_concepts}        & Explain concepts                             \\
                                & \texttt{ai\_understand\_codebase}     & Navigate, locate files, understand structure \\
      \midrule
      Eval                      & \texttt{ai\_critique\_output}         & Evaluate correctness, suggest improvements   \\
      \midrule
      \multirow{3}{*}{Setup}    & \texttt{ai\_setup\_environment}       & Configure env, install deps, build tools     \\
                                & \texttt{ai\_git\_operations}          & Git commands, branching, merging             \\
                                & \texttt{ai\_run\_or\_deploy}          & Run tests, start servers, deploy             \\
      \midrule
      \multirow{2}{*}{Converse} & \texttt{ai\_acknowledge}              & Acknowledge, confirm, greet; non-task input  \\
                                & \texttt{ai\_provide\_context}         & Share logs, terminal output, or context      \\
      \bottomrule
    \end{tabular}%
  }
  \caption{Behavior codebook (17 codes, 6 categories). The Plan, Code, Explain, and Eval categories are adopted from \citet{ma2026dspm}; we introduce Setup (environment, git, deployment) and Converse (acknowledgments, context sharing) to cover prompts that fell outside the original categories.}
  \label{tab:codebook}
\end{table}

\section{Case Study: Classroom Deployment}
\label{sec:case-study}

We deployed \textsc{RECAP} in a software engineering for machine learning course at an R1 University in the United States in Spring 2026.
In a two-week assignment, students extended two LLM-based features to Zulip,\footnote{\url{https://github.com/zulip/zulip}} an open-source team collaboration tool.
We present this deployment to demonstrate that the platform works at scale and to illustrate the kinds of analyses it enables.

\paragraph{Scale.}
\textsc{RECAP} captured data from 41 students who used GitHub Copilot: 29 produced chat data (2,034 prompts) and all 41 produced shadow git data (8,239 commits) across 406 work sessions.
The gap reflects students who used AI tools outside VS~Code's Copilot Chat (e.g., ChatGPT in a browser); the shadow git captures all code changes regardless of which AI tool was used.

\begin{figure}[t]
  \centering
  \includegraphics[width=\columnwidth]{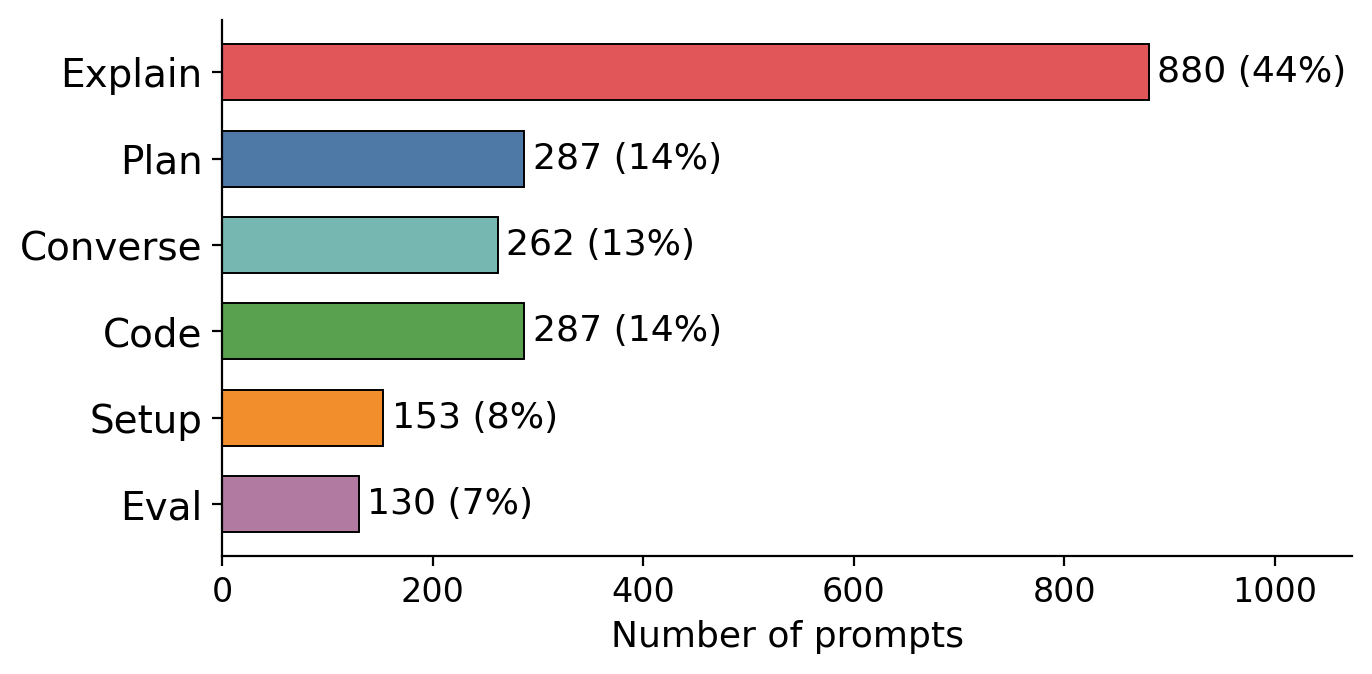}
  \caption{Distribution of prompt behavior categories; Explain dominates over Code.}
  \label{fig:behaviors}
\end{figure}

\begin{figure}[t]
  \centering
  \includegraphics[width=\columnwidth]{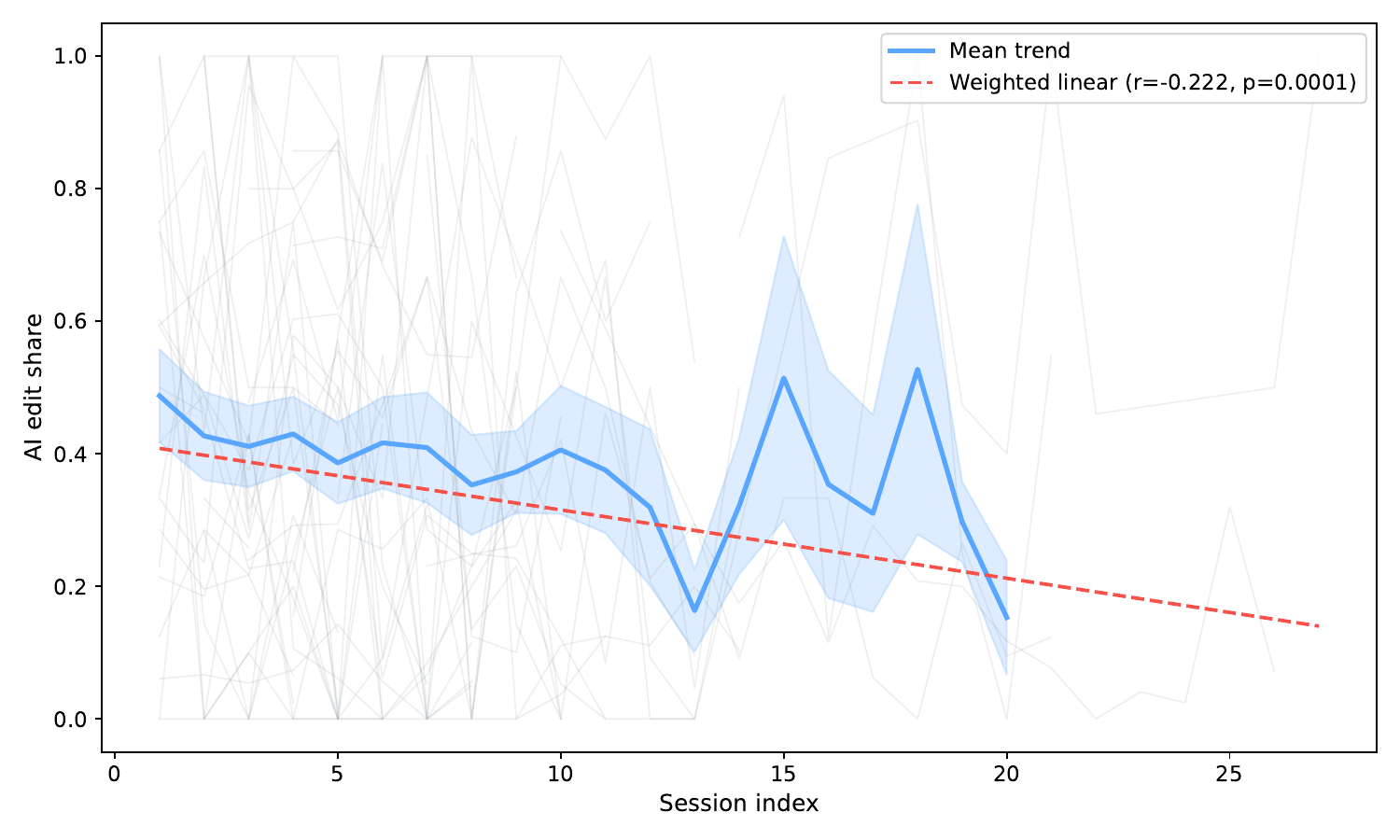}
  \caption{AI edit share across successive work sessions. Blue: mean trend $\pm$ SE. Red dashed: weighted linear fit ($r\!=\!{-}0.222$, $p\!<\!0.001$). Gray: individual student trajectories.}
  \label{fig:temporal}
\end{figure}

\paragraph{Behavioral overview.}
The analysis pipeline classified prompts into 6 categories (Figure~\ref{fig:behaviors}).\footnote{We exclude 32 trivial prompts (e.g., \texttt{yes}, \texttt{OK}, \texttt{hi}); 3 prompts ambiguous out of context (e.g., \texttt{I want A}) are classified as Other and not shown.}
\textbf{Explain} dominates (44\%, with \emph{explain error} alone at 29\%), followed by \textbf{Plan} (14\%), \textbf{Code} (14\%), \textbf{Converse} (13\%), \textbf{Setup} (8\%), and \textbf{Eval} (6\%), suggesting students may use AI more for comprehension than for generation.
AI edit share (fraction of edits attributed to AI) trends downward over successive sessions ($r\!=\!{-}0.222$, $p\!<\!0.001$, Figure~\ref{fig:temporal}), indicating that students' use of AI for code edits decreases over the course of the project.

\paragraph{Qualitative patterns from replay.}
Beyond aggregate statistics, the replay viewer surfaces interaction patterns that are difficult to recover from chat logs or git histories alone.
We highlight three examples from the deployment:

\emph{Error-pasting loop.}
One student spent 11 minutes cycling through the same \texttt{TypeError} three times.
The AI fixed each occurrence superficially, surfacing a new error that led back to the original.
The replay timeline makes this cycle visible: alternating prompt and edit markers with no forward progress in the diff view.
Appendix~\ref{app:behaviors} shows this pattern in the replay viewer (Figure~\ref{fig:error-loop}).

\emph{Cross-tool usage.}
After hours of failed attempts, a student turned to ChatGPT for an architectural suggestion and pasted it directly into Copilot Chat: ``This is what ChatGPT says and I think we should try and implement that.''
The AI generated edits across multiple files, but the approach still failed.
This pattern---using one AI for strategy and another for implementation---is only visible when chat content and code outcomes are linked.

\emph{Agentic generation and iterative refinement.}
A student prompted the Copilot agent with the full assignment spec for each of two features.
For Feature~1, the agent autonomously created multiple files across frontend and backend in its initial turns. Eight follow-up prompts, shifting from broad directives (``add UI button below drafts'') to precise bug reports (``full stop is also becoming part of the link''), brought the feature to completion.
The student repeated the same approach for Feature~2, pasting the spec and relying on agentic generation. However, the more complex task (backend, frontend, external API, database) led to cascading build and syntax errors that were not resolved as quickly.
In this case, the contrast suggests that the generate-then-debug workflow may succeed for simpler tasks yet struggle as complexity grows; replay surfaces the divergence.

\section{Related Work}

As AI coding tools evolve from inline auto-completion to long-horizon agents like GitHub Copilot, Cursor, and Claude Code, understanding developer–AI collaboration requires more than chat logs or code commits in isolation. Prior work has examined programming behavior, LLM usage, and AI-assisted development, but often under constrained settings or short tasks. RECAP builds on current literature while addressing gaps in ecological validity, task scale, and replayable analysis.

\paragraph{IDE Telemetry and Developer Workflows.}
Prior systems instrument developer workflows to study both professionals and students. Large-scale, opt-in naturalistic logging systems such as Blackbox \cite{Brown2014-vd} collected activity data from real users at scale. 
Efforts like ProgSnap2 aim to standardize log collection and analysis with shared schemas \cite{Price2020-ze}. Keystroke, snapshot, or IDE telemetry systems show that replays especially support inference over developer strategies \cite{Karol2025-cm}, tutoring conversations \cite{Yan2019-kd}, and self-regulation \cite{Xie2023-xi}. Some tools move beyond version control histories and also capture recordings with web browsers \cite{Pham2025-us}. The rich literature in program evolution research provides insights transferable to AI-assisted programming logging in that we need (1) fine-grained, diff-able histories beyond commits, (2) privacy-aware, large-scale capture in authentic environments, and (3) replay and inspection tools to analyze behaviors.

\paragraph{LLM Interactions and User Behaviors.}
LLM interaction research, especially with writing and general chat tasks, contributes methodologies for studying multi-turn human–AI collaboration~ \cite{Mysore2025-bg}. For example, CoAuthor \cite{Lee2022-ik} logged prompts, suggestions, and revisions for human–LLM co-writing. 
These works show that understanding LLM-assisted work needs linking prompts to downstream outcomes, which are evolving code states for programming. 
In programming-related benchmarks, DevGPT \cite{Xiao2024-fv} links shared ChatGPT conversations to downstream software artifacts, offering breadth but relying on self-selected data rather than continuous IDE instrumentation. 
StudentEval \cite{babe2024studenteval} captures prompt–model interactions from novices but does not provide naturalistic IDE traces of real projects. RealHumanEval \cite{Mozannar2024-wq} collect structured interactive traces but require participants to use constrained interfaces and short tasks.

\paragraph{Scalable Analysis of AI-Assisted Programming.}

Lab studies \cite{barke2023grounded, Bird2023-jv, Mozannar2022-ng} typically characterize interaction modes and user perceptions over minutes or hour-long tasks and focus on metrics such as completion time and code quality \cite{Ma2023-as}. 
In classroom studies, \citet{ma2026dspm} analyzes student–AI interactions in Colab notebooks, and CodeAid \cite{kazemitabaar2024codeaid} replaces the default IDE with a custom LLM assistant.
Prior works also analyze repository-level evolution under only Cursor use \cite{He2026-tp} or controlled agent conditions for GitHub Copilot and OpenHands agents \cite{Chen2025-ey}.

Various toolkits were developed to support logging of coding agent usage, such as Cursor's Agent Traces \cite{Cursor2026-ep} and Anthropic's 
Clio \cite{Tamkin2024-ub}. 
Editrail \cite{Zhang2026-je} records keystroke-level and browser-based AI interactions in GitHub Copilot, visualizing ``AI trails'' for instructors, but focuses on short tasks (150–300 lines, ~20 minutes). Hodoscope \cite{zhong2026hodoscope} analyzes agent trajectories on benchmarks, enabling large-scale behavioral pattern discovery. 
Many recent tools lack either chat prompt or agentic tool call data \cite{Basha2025-sl, Sergeyuk2026-rq, Park2025-uo, chi2025copilotarenaplatformcode}. SWE-chat \cite{Baumann2026-rw} provides a large-scale dataset with naturalistic prompts, agent responses, and code; however, it does not track code edits beyond commits.

While these studies provide valuable tools and behavioral characterizations, they are generally time-, platform-, and/or task-constrained. 
\textsc{RECAP} aims to address the gap by providing scalable IDE instrumentation, automatic prompt-to-edit linking, interactive replay, and automated behavioral analysis in a single open-source toolkit.

\section{Conclusion}

We presented \textsc{RECAP}, an open-source platform that captures AI chat sessions and fine-grained code edits inside VS~Code, merges them into an interactive session replay, and provides extensible analysis modules.
Deployed with 41~students on weeks-long projects, \textsc{RECAP} demonstrated its ability to capture agentic AI interactions at scale and surface candidate patterns --- error-pasting loops, cross-AI usage, and shifts in AI edit share --- that motivate further study with linked traces.
Future work includes extending capture to other AI assistants and IDEs such as Cursor, connecting behavioral patterns to learning outcomes in classrooms, and longitudinal analysis across developer and student populations.

\section*{Limitations}

\textsc{RECAP} currently captures data only inside VS~Code and is specific to GitHub Copilot. Extending to support other editors (e.g., Cursor) and agent types (e.g., Claude Code, Copilot CLI, Anthropic and OpenAI agents in VS~Code) is needed in future work.
The behavior classifier relies on a commercial LLM, introducing cost and potential inconsistency.
The ``External Source'' heuristic may produce false positives when developers type large code blocks manually.
Our case study covers a single course; generalization to professional developers requires further validation.

\section*{Ethics Statement}

All data collection was conducted under IRB approval.
Students were informed about data collection at course start, and participation in research analysis was voluntary.
Identifiers are SHA-256 hashed client-side, and data is stored in access-controlled cloud storage.
The Replay Viewer is for authorized researchers and instructors only; we do not release individual student data.

\section*{Acknowledgments}

This work is partially supported by the National Science Foundation (awards CNS-2213791 and 2414915), the Google Academic Research Award, and the Amazon AI Research Award.
We thank all participating students and instructors in our study, and all WInE and LearnLab members who provided feedback on this work.

\bibliography{custom,paperpile}

\appendix

\section{Prompt Embedding Visualization}
\label{app:embedding}

\begin{figure*}[t]
  \centering
  \includegraphics[width=\textwidth]{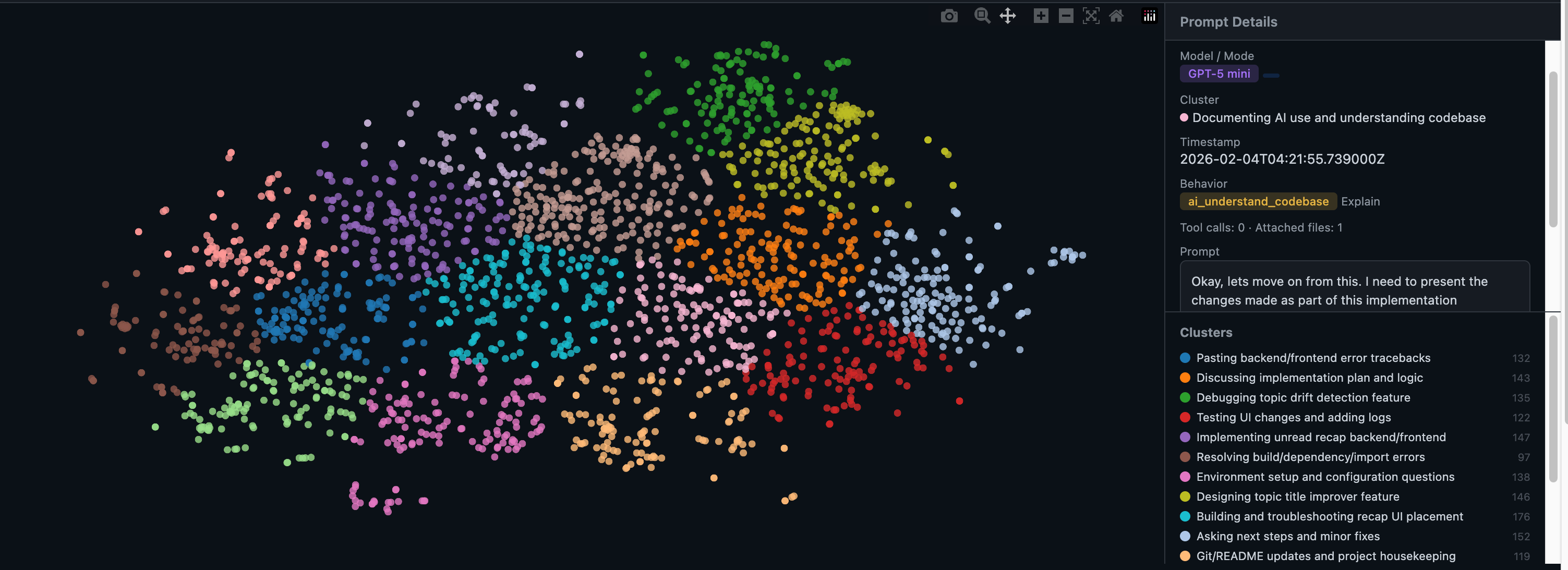}
  \caption{t-SNE projection of 2,034 student prompts colored by behavior category. Clusters reveal distinct interaction patterns (e.g., error debugging, endpoint testing, UI styling).}
  \label{fig:embedding}
\end{figure*}

Figure~\ref{fig:embedding} shows a t-SNE projection of all 2,034 prompts from the deployment, colored by behavior category.
Prompts with similar intent cluster together.
The visualization serves as an exploratory tool, and researchers can interactively color by student, model, time period, or cluster ID to discover patterns such as which students rely heavily on a single prompt type or how prompting strategies shift over the course of a project.

\section{Qualitative Behavior Example}
\label{app:behaviors}

This appendix provides a replay screenshot and additional detail for the \emph{error-pasting loop} pattern described in \S\ref{sec:case-study}.

Over an 11-minute span (prompts~17--23) the student pasted the same \texttt{TypeError: get\_unread\_messages\_recap() takes 1 positional argument but 2 were given} multiple times. Each AI response confidently identified a different cause and applied a corresponding edit to \texttt{views.py}. Each fix resolved the immediate symptom but reintroduced the original \texttt{TypeError} on the next request. Figure~\ref{fig:error-loop} shows three prompts and AI responses from the loop.

\newcommand{\imgcol}[2]{%
  \begin{minipage}[t]{0.3\textwidth}\centering
    \small\textbf{#1}\\[2pt]
    \includegraphics[width=\textwidth]{#2}
  \end{minipage}%
}

\begin{figure*}[t]
  \centering
  \imgcol{Prompt \#17}{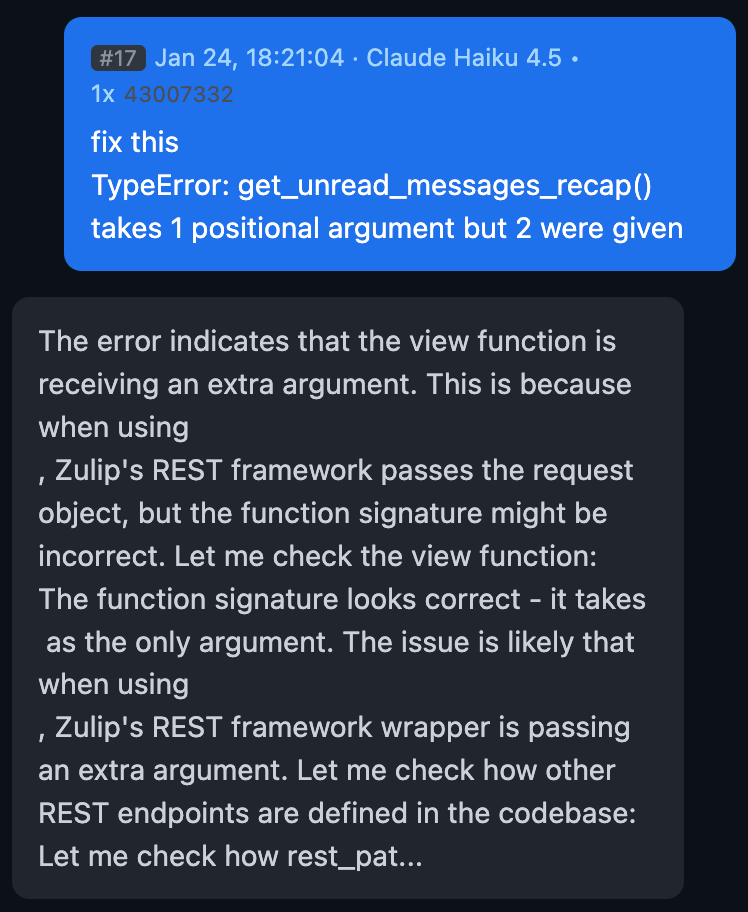}%
  \hfill\raisebox{-3cm}{\Large\textbf{\ldots}}\hfill
  \imgcol{Prompt \#20}{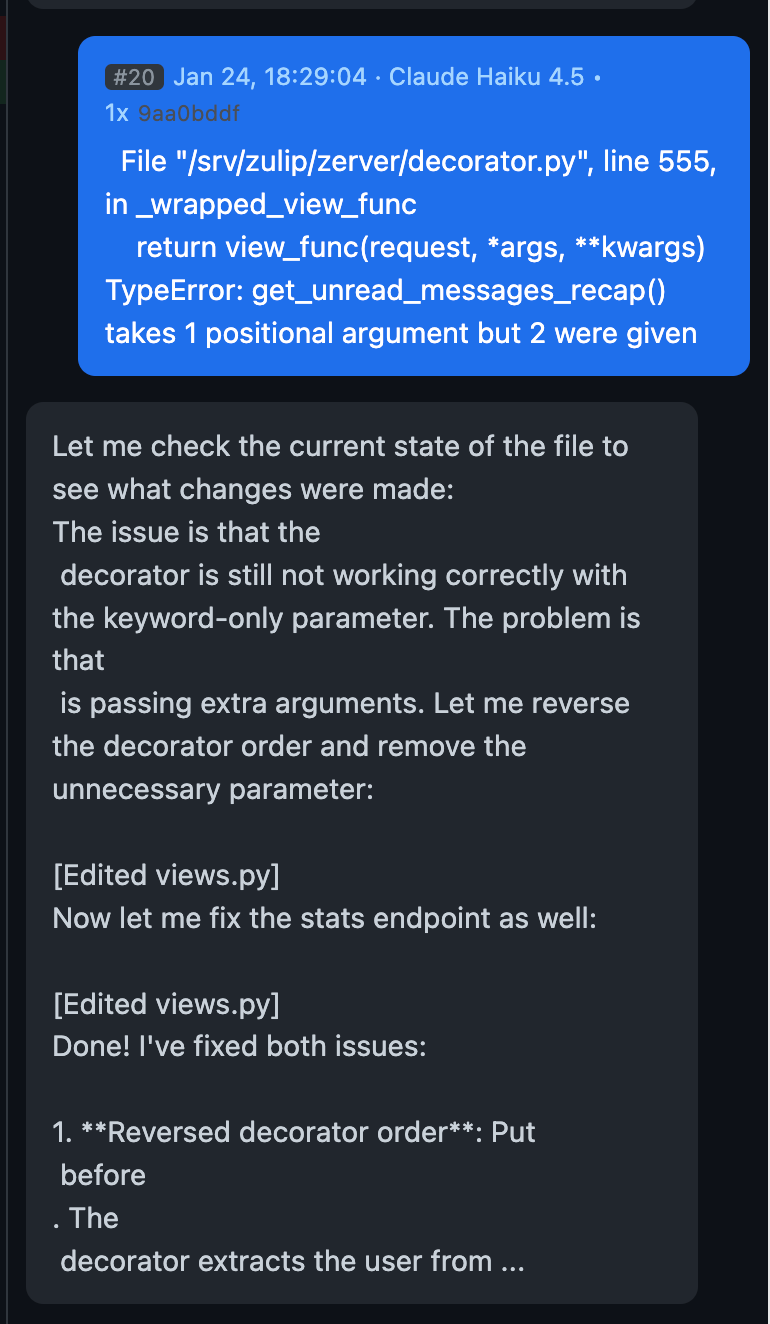}%
  \hfill\raisebox{-3cm}{\Large\textbf{\ldots}}\hfill
  \imgcol{Prompt \#23}{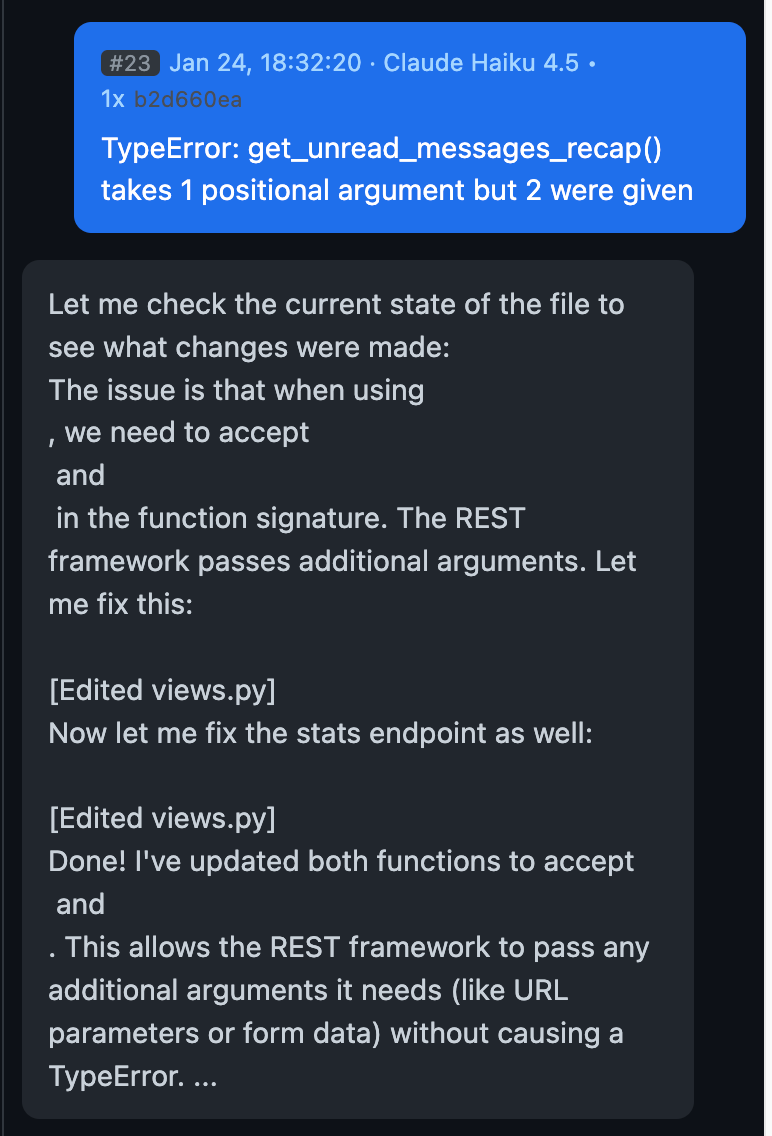}
  \caption{Error-pasting loop in the replay viewer. Over 11 minutes the student pastes the same \texttt{TypeError} into Copilot Chat three times (prompts 17, 20, 23; intervening prompts omitted with ``\ldots''). Each AI response confidently ``fixes'' the issue by editing \texttt{views.py}, but the fix introduces a different decorator error whose resolution brings back the original \texttt{TypeError}.}
  \label{fig:error-loop}
\end{figure*}

\end{document}

%% file: figures/architecture.tex
\begin{figure*}[t]
  \centering
  \resizebox{\textwidth}{!}{%
    \begin{tikzpicture}[
        >=Stealth,
        box/.style={draw, rounded corners=3pt, minimum height=1.1cm, align=center, font=\small},
        grp/.style={draw, rounded corners=5pt, inner sep=6pt},
        arrow/.style={->, very thick, color=gray!70},
      ]

      \node[box, fill=green!15, text width=1.6cm] (chat) {Chat\\Watcher};
      \node[box, fill=green!15, text width=1.6cm, right=0.3cm of chat] (git) {Shadow\\Git};
      \node[box, fill=green!15, text width=1.6cm, right=0.3cm of git] (log) {Log\\Tailer};

      \begin{scope}[on background layer]
        \node[grp, fill=green!6, fit=(chat)(git)(log),
          label={[font=\footnotesize\bfseries]above:VS Code Extension}] (collect) {};
      \end{scope}

      \node[box, fill=blue!12, text width=1.6cm, right=1.2cm of log] (cloud) {Cloud\\Storage};

      \node[box, fill=orange!15, text width=1.8cm, right=1.2cm of cloud] (timeline) {Timeline\\Builder};
      \node[box, fill=orange!15, text width=1.8cm, right=0.3cm of timeline] (replay) {Replay\\Viewer};

      \begin{scope}[on background layer]
        \node[grp, fill=orange!6, fit=(timeline)(replay),
          label={[font=\footnotesize\bfseries]above:Replay Platform}] (platform) {};
      \end{scope}

      \node[box, fill=red!10, text width=2.2cm] (behav) at ($(cloud.south)+(-3.9cm,-1.6cm)$) {Behavior\\Classification};
      \node[box, fill=red!10, text width=2.2cm, right=0.4cm of behav] (reliance) {AI\\Reliance};
      \node[box, fill=red!10, text width=2.2cm, right=0.4cm of reliance] (embed) {Prompt\\Embedding};
      \node[box, fill=red!10, text width=1.6cm, right=0.4cm of embed, font=\small\itshape] (custom) {\ldots};

      \begin{scope}[on background layer]
        \node[grp, fill=red!4, fit=(behav)(reliance)(embed)(custom),
          label={[font=\footnotesize\bfseries]above:Extensible Analyses}] (analyses) {};
      \end{scope}

      \draw[arrow] (collect.east) -- (cloud.west);
      \draw[arrow] (cloud.east) -- (platform.west);
      \draw[arrow] ($(analyses.west)+(-1.0cm,0)$) -- (analyses.west);

    \end{tikzpicture}
  }%
  \caption{\textbf{\textsc{RECAP} system architecture.} The VS~Code extension captures chat sessions and code edits in parallel and uploads them to cloud storage. The replay platform merges them into a unified timeline for interactive replay. Extensible Analyses are example modules built on the timeline; researchers can plug in their own without modifying the capture layer.}
  \label{fig:architecture}
\end{figure*}